\documentclass[aps,superscriptaddress,preprint,showpacs]{revtex4}
\usepackage{bm}
\usepackage{dcolumn}
\usepackage{epsfig}

\begin{document} 

\title{Holographic Principle and Quantum Physics}

\author{Zolt$\acute{{\rm a}}$n Batiz 
\footnote{zoltan@cftp.ist.utl.pt} 
and Bhag C. Chauhan 
\footnote{chauhan@cftp.ist.utl.pt}
}
\affiliation{
\it Centro de F\'{i}sica Te\'{o}rica das Part\'{i}culas (CFTP)\\
Departmento de Fisica, Instituto Superior T\'{e}cnico\\
Av. Rovisco Pais, 1049-001, Lisboa-PORTUGAL}

\date{\today}

\begin{abstract}
The concept of holography has lured philosophers of science for decades, 
 and is becoming more and more popular in several front areas of science,
e.g. in the physics of black holes.
In this paper we try to understand things as if the visible universe were a 
reading of a lower dimensional hologram generated in hyperspace. 
We performed the whole process of creating and reading holograms of a point 
particle in a virtual space by using computer simulations. 
We claim that the fuzziness in quantum mechanics, in statistical physics and 
thermodynamics is due to the fact that we do not see the real image of the 
object, but a holographic projection of it. 
We found that the projection of a point particle is a de Broglie-type wave.
This indicates that holography could be the origin of the wave nature of a 
particle. We have also noted that one cannot stabilize the noise 
(or fuzziness) in terms of the integration grid-points of the hologram, it 
means that one needs to give the grid-points a physical significance. So we 
futher claim that the space is quantized, which supports the basic assumption 
of quantum gravity. 
Our study in the paper, although is more qualitative, yet gives a smoking gun 
hint of a holographic basis of the physical reality.  
\end{abstract}


\maketitle

\newpage
\section{Introduction}
\label{sec:introduction}
The theory of holography was first developed by Hungarian scientist 
Dennis Gabor around 1947-48 while working to improve the resolution of an 
electron microscope \cite{gabor}. 
The first holograms were of poor quality, but the principle was good.  
According to the principle of holography, a detailed three dimensional image 
of an object can be recorded in a two dimensional photographic film 
and the image can be reproduced back in a three dimensional space. 
The complex patterned information stored in the film is called 'hologram' 
(see in Fig. 1). 
The holograms have a strange feature, unlike the conventional photographic 
film, once you cut a hologram into pieces each piece is capable of 
reconstructing the entire image, although with lesser and varying resolutions.
  
\begin{figure}[hbt]
\centerline{\epsfig{file=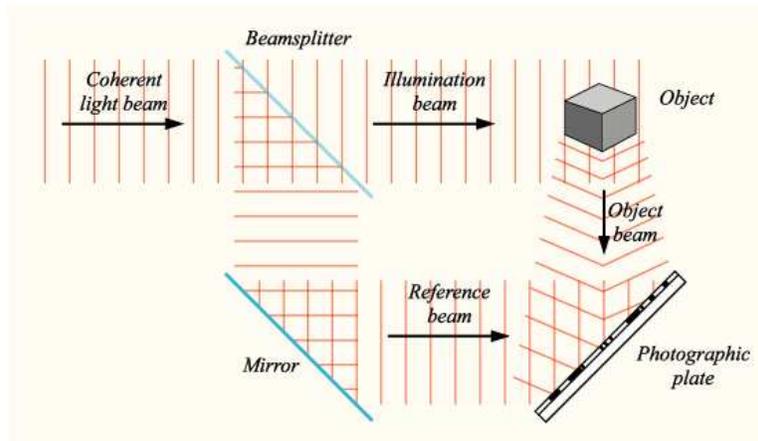,width=4.0in}}
\caption{Image taken from \cite{wiki}}
\label{fig2}
\end{figure}

The holographic concept has lured philosophers of science for decades 
\cite{bohm, talbot}, and is becoming more and more popular in several front
areas of science; attracting the researchers of cosmology, astrophysics, 
extra-dimensions, string theory, nuclear and particle physics and 
neurology \cite{pribram} etc... One can find hundreds of papers in the 
internet discussing the relevance of holography in these fields.  
Some articles claim that our visible and highly complex universe is 
actually a hologram of a higher dimensional but simpler reality. 
The holographic principle is now widely being used to 
relate seemingly unrelated things, like quantum mechanics and gravity 
\cite{susskind}.
Theoretical results about black holes suggest that the universe could be 
like a gigantic hologram \cite{jacob}. 
So our seemingly three-dimensional universe could be completely equivalent to 
alternative quantum fields and physical laws "painted" on a distant, 
vast surface.

On the same lines, in our paper we try to understand things as if the
visible universe were a reading of a lower dimensional hologram 
generated in hyperspace. 
%
Our procedure goes as follows. 
To start with, we perform the whole process of creating and reading  
holograms in a virtual space by using computer simulations. 
For simplicity we consider a point object with no geometrical size
at rest and at uniform motion.
First, we create the hologram of the point object using a reference beam and 
an object beam.
Second, we read the created hologram by illuminating it with a 
suitable reference beam. 

Ideally, the reconstruction procedure 
should give back the original point object in our normal physical 
space, but we know that no optical procedure gives a 
strictly point-like image of a point-like object. 
Assuming holography as the basis of physical reality, we claim 
that some dynamics seen in the observable world 
(such as fuzziness in quantum mechanics, in statistical physics and 
thermodynamics) can come from such a procedure. This fuzziness is basically a 
noise that has a dynamics of its own. In this work we will try to understand
this in detail.

In the first section \ref{S:form} we describe the creation 
and reading of hologram in usual space and hyperspace. 
In the second section \ref{S:stationarypoint} we study the image created by 
the holographic mapping of a single stationary point, and 
then reading the hologram. 
In the third \ref{S:movingpoint} section we repeat this 
procedure for a linearly and uniformly moving point. 
Both \ref{S:stationarypoint} and \ref{S:movingpoint}
will suggest that the image of a point-like particle is a wave that 
is described by a de Broglie-type of relation \cite{de-Broglie}. 
This supports our assumption of a holographic base of physical reality.  
We have also noted that one cannot stabilize the noise (or fuzziness) in terms 
of the integration grid-points of the hologram screen, it means that one needs 
to give the grid-points a physical significance. So we futher claim that the 
space is quantized, which concurs with the basic assumption of quantum 
gravity. 

In section \ref{S:conclusions} we present the discussion and conclusions.
 
\section{Holographic mapping in higher dimension and reading of the hologram}
\label{S:form}
In this section we first describe the creation  
of holograms in usual space and hyperspace. We create a two-dimensional
hologram (\ref{S2:form}).
Next, we describe the reconstruction of the hologram in the usual 
three-dimensional space (\ref{S1:form}). It must be noted that 
the holograms are not necessarily created by light, but can be formed in the
 presence of any wave action.
In principle they can be created and
read with different kind of waves, such as scalar, vector 
(electromagnetic, acoustic), tensor (gravitational), 
and the calculation principles are basically the same \cite{BW}. 

Since the motivation is to understand the usual three-dimansional world 
reality as a holographic projection of some higher dimensional reality, we 
realize that we need atleast five dimensions (3+2) to create the usual 
three-dimensional image (the objects we perceive through our senses).
The extra two dimensions are required for the reference and object 
beams to propagate, since we do not see them. 
%
%
\subsection{Creating hologram} 
\label{S2:form}
The hologram itself is the picture of the interference pattern of a 
beam that comes from an object we want to map and a reference beam.
For simplicity we assume a scalar beam, but the result is the 
same for other waves as well \cite{BW},  since, 
according to this reference, the interference pattern does not 
depend on the spinorial structure of the wave (like scalar, vector, etc.). 
The hologram as well as its reading are interference patterns, so 
this independence is understandable.
The hologram is produced, 
as seen in Fig. 1 and 2, by splitting a single beam into 
two pieces: 
one is shed on the object and the other directly on the photographic plaque.
In our case we assume a five-dimensional space with the holographic screen 
lying on the $(x,y)$ plane by construction. We take a beam that hits the 
object propagates in the fifth dimension and a reference beam that propagates 
in the $(y,x_4)$ plane. And there is an angle $\psi_r$ between the 
beam itself and the $x_4$ axis. 
We didn't take it in the $(x_4,x_5)$ plane
because in such case the reference wave that hits the screen remains always 
normal to it. 
We assume that the phase of the 
reference beam is $\phi_{r_i}$ at the point $(y,x_4)=(0,0)$.  
The wavelength of the radiation is $\lambda=2 \pi/k$ 
and $k$ is the wave number.
Therefore the phase of the reference beam 
at an arbitrary site is $\phi_r=\phi_{r_i}+kx_4\cos(\psi_r)-k y \sin(\psi_r)$. 
On the screen $x_4=0$, so $\phi_r=\phi_{r_i}-k y \sin(\psi_r)$.

This wave has an amplitude $E_{r0}$, so its value is 

\begin{equation}
{E}_{r}=E_{r0}\exp[i(\phi_{r_i}-ky \sin(\psi_r))].
\label{1aeq1a}
\end{equation}

\begin{figure}[hbt]
\centerline{\epsfig{file=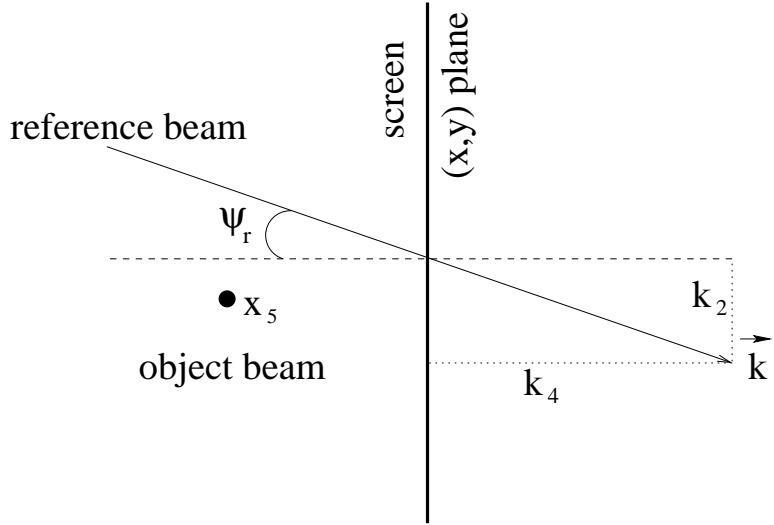,width=4.0in}}
\caption{A schematic diagram showing the reference beam propagation in 
($y,x_4$) plane into the photographic film on ($x,y$) plane. 
The dot represents the object beam propagating in the fifth dimension.}
\label{fig1}
\end{figure}

Likewise the phase of the beam that is reflected from the object is
$k(x_5+r)$, where $r$ is the distance between the object 
and an arbitrary point on the screen where we compute the phase.  
The reflected wave therefore is

\begin{equation}
{E}_{o}=E_{o0} e^{i (\phi_{r_o}+k (x_5+r))}/r^{(d-1)/2}, 
\label{1aeq1}
\end{equation}

where $E_{o0}$ is a constant proportional to the square root 
of the surface of our object, the square root of its reflectivity ($R$) and 
the original field strength of the object beam. Here $d$ 
is the dimensionality
of our space. However, $\phi_{r_o}$ is the initial phase of the wave which 
hits the object. These two fields can be added together and squared, and 
then we have the interference picture that is the hologram.

How do we determine $E_{o0}$? The energy that hits the object is 
$S_d E_{o0}^2/2$, where $S_d$ is the ``cross section'' of the object 
in $d$ dimensions. Thus, in two dimensions it is $2r_o$ (where $r_o$ is the 
radius of the ``spherical'' object) in three dimensions it is $\pi {r_o}^2$, 
and by generalization, for $d$ dimensions
it is $V_{d-1}$, where $V_d$ is the volume of the $d$ dimensional sphere.
The energy that comes out ($I_{o}$) is $R^2$ times the one that goes in 
($I_{i}$). 
 The energy is spread out on the whole ``solid angle'' $\Omega_d$.
$\Omega_d$ is determined from the following identities:

\begin{eqnarray}
\int d^d x \exp{(-ax^2)}=&&\Omega_d \int_0^{\infty}\exp{(-ar^2)}
r^{d-1}dr \\ \nonumber
\int d^d x \exp{(-ax^2)}=&&\left(\frac{\pi}{a}\right)^{d/2},
\end{eqnarray}

and it is found to be 

\begin{eqnarray}
\Omega_d=\frac{2 \pi^{d/2}}{\Gamma(\frac{d}{2})},
\end{eqnarray}

where $\Gamma$ is the Euler function. 
The $d$ dimensional volume is $\Omega_d r^d/d$. The energies going 
in and coming out are given as $I_{o}= \Omega_d r^{d-1} E_{o}^2/2 $ 
and $I_{i}= S_d E_{i}^2/2$, respectively. As discussed 
before they are related as

\begin{eqnarray}
I_{o}=R^2 I_i, \nonumber
\end{eqnarray} 

such that

\begin{eqnarray}
|E_{o}|=|E_{i}| \sqrt{\frac{S_d R^2}{r^{d+1}}},
\end{eqnarray} 
where $E_{i}$ and $E_{o}$ are in-going and out-going waves, respectively.
In our calculations $E_{i}$ and $S_{d}$ are the numerical inputs, which we
choose so as to get the clear wave pattern with minimum noise. 
 
\subsection{Reading hologram} 
\label{S1:form}
The hologram is read once one sheds a wave on it and, the direction and the 
frequency of this beam should be same as those of the reference beam 
that was used to create it. In order to read hologram we will 
disgress from the way we created it. We will read it in 4D space, since we 
don't need object beam (the fifth dimension), however our 
reading beam is the same as the original reference beam that created the 
hologram. 

In order to determine the image generated by a hologram when we 
shed some wave onto it, one must know the reflected fields at any given point. 
The square of those fields is proportional to the intensity of the 
reflected radiation, and knowing that, we can have an analytical 
description of the generated image.
If some wave is reflected from a surface (such as a hologram), we can
compute the reflected fields on the surface and we can examine 
how those fields propagate. First we consider how static fields are 
determined from known boundary conditions and after that we 
extend the calculation for wave fields. 

The Green function $G({\vec r})$ of a static field is defined as:

\begin{equation}
{\nabla}^2 G({\vec r})=-\delta({\vec r}).
\label{1eq1}
\end{equation}
The field at any given point can be calculated as: 
\begin{eqnarray}
\varphi({\vec r}^{\prime})=\int_V d^3 {\vec r} 
\varphi({\vec r})\delta({\vec r},{\vec r}^{\prime})=
-\int_V d^3 {\vec r} \varphi({\vec r}) 
{\nabla}^2 G({\vec r},{\vec r}^{\prime}),
\label{1eq2}
\end{eqnarray}

where the derivative ${\nabla}$ is related to the variable ${\vec r}$.
Here $\vec r$ and $\vec r^{'}$ are the distance vectors of the fields 
${\varphi}$ on the screen and on a given point.

After applying the following identities:

\begin{eqnarray}
\varphi({\vec r}) 
{\nabla}^2 G({\vec r},{\vec r}^{\prime})=&&{\vec {\nabla}}
\left[ \varphi({\vec r}) {\vec {\nabla}}G({\vec r},{\vec r}^{\prime}) \right]
-({\vec {\nabla}}G({\vec r},{\vec r}^{\prime}))
({\vec {\nabla}}\varphi({\vec r})) \, ,\nonumber \\ 
({\vec {\nabla}}G({\vec r},{\vec r}^{\prime}))
({\vec {\nabla}}\varphi({\vec r}))=&&
{\vec {\nabla}}(G({\vec r},{\vec r}^{\prime}) 
\nabla^2 \varphi({\vec r}))\, ,
\label{1eq2a}
\end{eqnarray} 

and making use of the Laplace equation:
\begin{eqnarray}
\nabla^2 \varphi({\vec r}))=-\rho({\vec r})\, ,
\label{1eq3}
\end{eqnarray}
the field in any point can be calculated as
\begin{eqnarray}
\varphi({\vec r}^{\prime})=\int_V d^3 {\vec r} 
G({\vec r},{\vec r}^{\prime})\rho({\vec r}) + \int_S d^2 {\vec r}
G({\vec r},{\vec r}^{\prime}) {{\nabla}_n}\varphi({\vec r})-
\int_S d^2 {\vec r}
\varphi({\vec r}){{\nabla}_n}G({\vec r},{\vec r}^{\prime}))\, ,
\end{eqnarray}
where ${{\nabla}_n}$ is the component of the derivative that is perpendicular
to the surface.
 
The first term refers to the sources and we assume that there 
are no sources in the part of space we examine. The other two terms are 
the so-called surface terms. Whenever the first surface term vanishes
(and the Green's function must be chosen accordingly, so that it 
vanishes on the surface) we must know the value 
of the field on the surface, and we are said to use the Dirichlet conditions.
If we know only the derivatives of the fields on the surface,
we must require that the normal derivative of the Green's function 
vanishes on the surface, we are said to make use of a 
Neumann Green's function. Note that any of these conditions can be met at 
any time (although not both at the same time) because Eq. (\ref{1eq1})
does not completely fix the Green's function, so 
we might add any term whose Laplacian is zero 
(in the region of space we are interested in) in such a 
way that the new Green's function satisfies either 
one of the two conditions. Because we can calculate the fields at the surface,
we will use a Dirichlet Green's function, 
so our field at any given point is expressed as follows:
\begin{eqnarray}
\varphi({\vec r}^{\prime})=-\int_S d^2 {\vec r}
\varphi({\vec r}){{\nabla}_n}G({\vec r},{\vec r}^{\prime}))\, .
\label{1eq3a}
\end{eqnarray}

From the image solution for the auxiliary electrostatic problem, the Green 
function for Dirichlet conditions can be calculated.
We need to know the Dirichlet Green's function 
on a plane. First we define some new variables ${\vec r}_{1}$ and 
${\vec r}_{2}$ as 
${\vec r}_{1,2}=(x-x^{\prime}) {\vec e}_1+
(y-y^{\prime}){\vec e}_2+(z \mp z^{\prime}){\vec e}_3$, 
(in terms of our orthonormal basis ${\vec e}_1$, ${\vec e}_2$, ${\vec e}_3$),
$r_{1,2}=\sqrt{({\vec r}_{1,2} \cdot {\vec r}_{1,2})}$.
In these terms, the Dirichlet Green's function is given in \cite{eyges} as:  
\begin{equation}
{\tilde G({\vec r},{\vec r}^{\prime})}=\frac{1}{4 \pi}
\left( \frac{1}{r_1}-\frac{1}{r_2} \right)\, .
\label{1eq4}
\end{equation}

If instead of static field we have wave fields,
this Green's function is replaced with 
\begin{equation}
G({\vec r},{\vec r}^{\prime};t)=\frac{1}{4 \pi}
\left( \frac{\delta(t-r_1/c)}{r_1}-\frac{\delta(t-r_2/c)}{r_2} \right)\, ,
\label{1eq5}
\end{equation}
where $c$ is the phase velocity of our wave. Since we consider 
only one frequency ($\omega$), we only need 
the Fourier transform of this Green's function, which is:
\begin{equation}
G({\vec r},{\vec r}^{\prime};\omega)=\frac{1}{4 \pi}
\left( \frac{\exp{(-ikr_1)}}{r_1}-\frac{\exp{(-ikr_2)}}{r_2} \right)\, ,
\label{1eq6}
\end{equation}
with $k=\omega/c$ as the wave number. We substitute 
this result into Eq. (\ref{1eq3a}).
How can we justify this substitution, since Eq. (\ref{1eq3a}) 
has been derived based on the assumption that the fields are static?
We know that if there is a wave field, Eq. (\ref{1eq3}) is replaced with
\begin{equation}
\left( {\nabla}^2 -\frac{\partial^2}{\partial t^2}\right)
\varphi({\vec r})=-\rho({\vec r}).
\label{1eq7}
\end{equation}

If we only consider one frequency and retarded waves only,
our field can be expressed as 
\begin{equation}
\varphi({\vec r},t)=\varphi({\vec r},t=0)\exp{[-i \omega(t-l/c)]},
\label{1eq8}
\end{equation}
where $l$ is the distance between the source and observer.
Substituting this into Eq. (\ref{1eq7}) and dividing the resulting equation by
$\exp{(-i \omega(t-l/c))}$ we obtain Eq. (\ref{1eq1}). 
So if we work with the time Fourier transforms 
of the wave fields and Green's functions and assume only one frequency,
we are able to make use of the static formulation 
of the problem using the Fourier transform of the 
Green's function we have just given in Eq. (\ref{1eq6}).
The normal derivative of this Green's function on the 
surface defined by the hologram is:
\begin{equation}
{\nabla}_n G({\vec r},{\vec r}^{\prime})=
-\frac{\partial}{\partial z}G({\vec r},{\vec r}^{\prime})\vert_{z=0},
\label{1eq9}
\end{equation}
which, because we assume 
much smaller wave lengths than the distances $r_{1,2}$, we approximate as:
\begin{eqnarray} 
{\nabla}_n G({\vec r},{\vec r}^{\prime})=&&  
\frac{1}{4 \pi}(-2ik) \left( \frac{(z-z^{\prime})\exp{(-ikr_1)}}{r_1}- 
\frac{(z+z^{\prime})\exp{(-ikr_2)}}{r_2} \right)\mid_{z=0} \\ \nonumber
=&&
\frac{ikz^{\prime}}{2 \pi} \frac{exp{(-ikr^{\prime})}}{r^{\prime}},
\label{1eq10}
\end{eqnarray}
and this we are able to substitute into Eq. (\ref{1eq3a}).
If the dimensionality were different (but greater than three), 
Eq. (\ref{1eq10}) would be modified in the following way:
\begin{eqnarray} 
{\nabla}_n G({\vec r},{\vec r}^{\prime})=
\beta kz^{\prime} \frac{exp{(-ikr^{\prime})}}{{r^{\prime}}^{d-2}},
\label{1eq10a}
\end{eqnarray}

where $d$ is the dimensionality of the space and $\beta$ 
is an irrelevant constant we do not even bother to give.
Whenever we are reading the hologram, the only relevant difference that 
comes from this formula is the phase given by the exponential, 
all the rest would be irrelevant. Therefore it does not matter whether we read 
our hologram in a three or four dimensional space. 
This we also confirmed by our numerical calculations. 

Now the only thing missing from the picture is the reflected 
field at any given point of the hologram.
The phase and amplitude of the reflected wave
depends on the phase and amplitude of the
reading wave:

\begin{eqnarray} 
{\vec E}_{rf}=|R_h|{\vec E}_{rd}\, ,
\label{1eq13}
\end{eqnarray}

where ${\vec E}_{rf}$ is the reflected wave, $R_h$ is the reflectivity of the 
hologram (it is the hologram data file generated in the previous step) 
and ${\vec E}_{rd}$ is the reading wave (exactly same as the reference beam in 
eq. \ref{1aeq1a}).

\begin{eqnarray} 
{\vec E}_{rd}={\hat x} E_{r0} \exp\left (-i k y\sin(\psi_r)+ 
i\phi_{r_i}\right), 
\label{1eq14}
\end{eqnarray}

where $E_{r0}$ is its amplitude, $\phi_{r_i}$ is its phase at the point 
$(y,x_4)=(0,0)$ and $\psi_r$ is its angle of incidence. By construction, 
the reading beam is the original reference wave (\ref{1aeq1}) that 
propagates in the $(y,x_4)$ plane. 
There is a phase shift of $\pi$ after the reflection.

We incorporate Eqs. (\ref{1eq3a}), (\ref{1eq10}), (\ref{1eq13}) 
and (\ref{1eq14}) in a numerical code
to compute the reflected 
fields (and therefore the intensities) at any given point of 
the space. Therefore, we obtained the image from the hologram we were reading.

\section{Images of a stationary point}
\label{S:stationarypoint}
In this section we describe the image created by reading a 
hologram of a stationary point.
Like any optical procedure, this process too will give 
us a blurry picture instead of a single point. We will study the 
dynamics of this blurriness and try to read some physics into it.

The arguments of this section are not nearly as rigorous 
as those of the next one,
they are actually some kind of hand-waving arguments that 
only suggest some possible conclusions rather than prove any. However, 
this is indicative for the direction of the next section that is 
going to be somewhat more rigorous than the present one.

We consider a point-like object in a five-dimensional space.
Why do we  need five dimensions? As said before, we consider a reference beam 
and an object beam that are not visible, therefore we need 
some extra dimensions. Since these two beams should 
propagate in different directions, the number of the extra 
dimensions is two. 
We then numerically create its hologram that we will 
read in our usual three-dimensional space, 
but the beam shed on our hologram will not be part of our 
three-dimensional space, since we cannot see them.

Now we assume that the point-like object is separated by the 
center of our rectangular 
screen (whose picture is the actual hologram) 
by a distance $D$, the sides of the screen are equal 
and their size is $d$,
and that the line of separation is perpendicular to the screen.
For the numerical integration that is involved in the 
reading of the hologram, we divided the screen into $40^2$ equal regions.

\begin{figure}[hbt]
\centerline{\epsfig{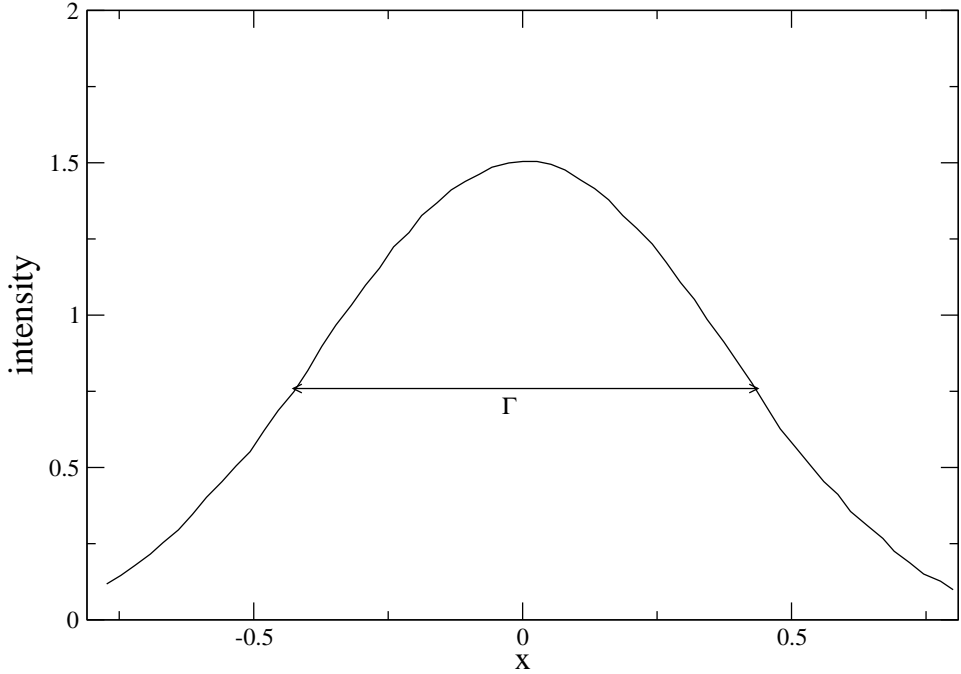}}
\caption{The image for $D=5$ units, $d=15$ units
and $\lambda=0.5 $ units. We found that 
$\Gamma=0.83$ units.}
\label{fig2a}
\end{figure}

First, we consider the case when the image of our point resembles a Gaussian 
distribution. This is the case, when, for example $D=5$ units, $d=15$ units
and $\lambda=0.5 $ units which 
we show also graphically in Fig. \ref{fig2a}. 

\vspace*{1.5cm}

\begin{table}
\begin{tabular}{|c|c|c|c|c|} \hline 
$D$  & 5  & 7.5  & 10 & 12   \\ \hline
$\Gamma$ & 0.83 & 0.64 & 0.40 & 0.30  \\
  \hline
\end{tabular}  
\caption{\it  Empirical study (particle at rest) of the parameters: $\Gamma$ and $D$}
\label{tab1}
\end{table}

$\Gamma$ is the width of the Gaussian curve.
From table I it appears that
$\Gamma \sim 1/D$. Gaussian curves are obtained if $\lambda \ll d$, 
$\lambda \ll D$ and $d > D$. 

\begin{table}
\begin{tabular}{|c|c|c|c|c|c|} \hline 
$D$  & 50  & 62.5  & 75  & 87.5  & 100  \\ \hline
$\Lambda$ & 1.0 & 1.3 & 1.6 & 1.9 & 2.2  \\
  \hline
\end{tabular}  
\caption{\it  Empirical study (particle at rest) of the parameters: $\Lambda$ and $ D$}
\label{tab2}
\end{table}

\begin{figure}[hbt]
\centerline{\epsfig{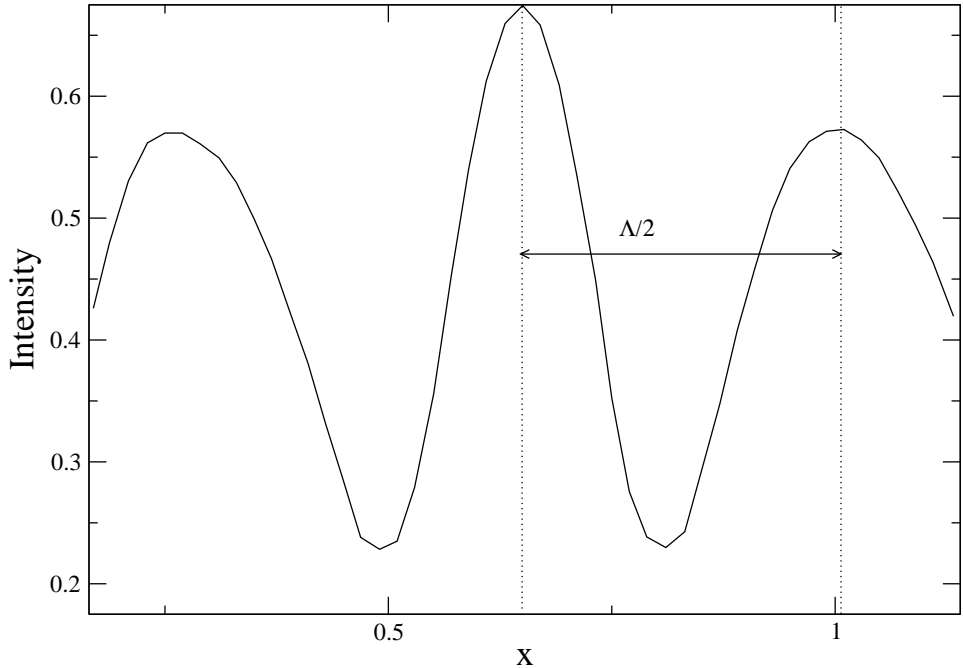}}
\caption{Holographic projection of a point particle at rest.
Here $D=75$ units, $d=0.4$ units
and $\lambda=0.005 $ units.}
\label{fig3}
\end{figure}

Another feature we obtained from our procedure is a wave like pattern.
This is the case, if $D=100$ units, $d=0.4$ units 
and $\lambda=5 \times 10^{-3}$ units.
Here $\Lambda$ is the wavelength of the pattern obtained.  
The specimen figure is shown in Fig. \ref{fig3}.
The table II shows that $\Lambda \sim D$. 

\vspace*{1.5cm}

Now here comes our hand-waving and very qualitative argument. 
From thermal physics we know that the width of the Gaussian distribution 
is proportional to the root mean square of the momentum, $p_{rms}$.
Therefore one can say that $\Lambda \sim D \sim 1/\sqrt{p^2}$, so $\Lambda
 \sim 1/p_{rms}$.
This is a de Broglie type relation, where the wavelength is inversely 
proportional with the momentum of the particle.
In our calculation we encountered a difficulty, however: the result cannot 
be stabilized in terms of the gridpoints. For a different number of gridpoints 
we are able to get the same patterns and the same proportionality relations,
but the graphs are different. This is understandable since the 
signal we investigate is basically a noise. 
The only way out is to suppose that space is quantized. 
Therefore not only statistical physics and quantum physics may be dependent, 
but qauntization of space could also be related to these two.

\section{Images of a Moving Point}
\label{S:movingpoint}
\begin{figure}[hbt]
\centerline{\epsfig{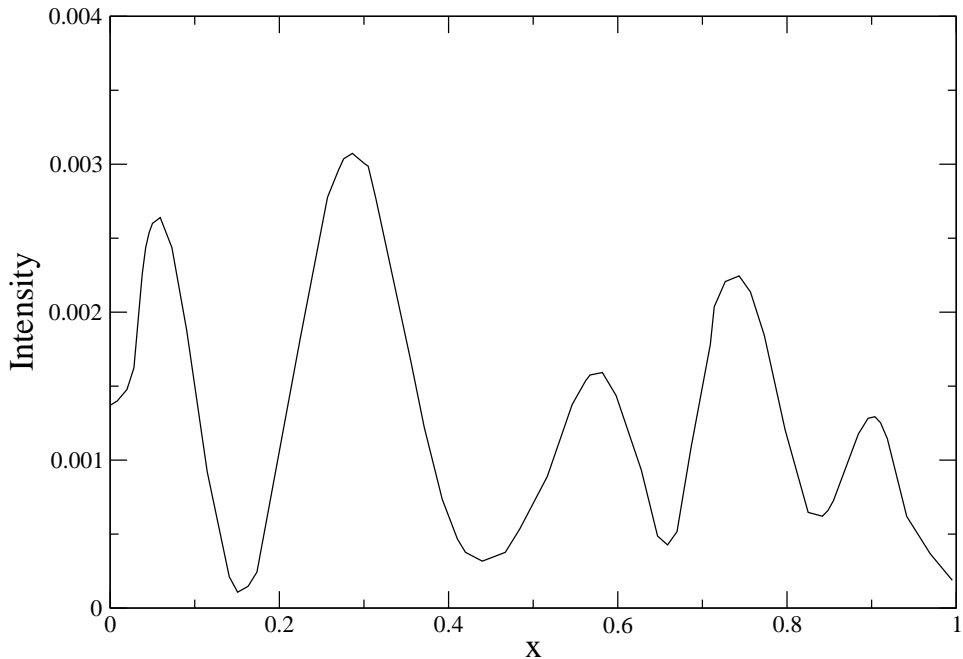}}
\caption{Holographic projection of a moving point particle.
Here $D=40$ units, $d=1.5$ units
and $\lambda=0.005 $ units.}
\label{fig4}
\end{figure}

A more elegant and compelling argument is the holographic 
mapping of a moving particle. We imagine that the hologram 
is taken during a finite interval of time; meaning that 
in every instant there is a snapshot, and these 
photographs are superposed, the resultant 
being the final hologram. The hologram then is read 
as we described in the previous sections. 
The emerging pattern in this situation is again a wave (see Fig. \ref{fig4}).
The only difference from the former calculations is that on the RHS
of Eq. (\ref{1aeq1}) the exponent $\phi_{r0}+kr$, which is the eikonal 
function of the wave field is replaced with the full eikonal. 
We do this because the moving point is also a dynamical system.
Therefore this exponent becomes $\phi_{r0}+kr+c{\vec P}.{\vec{\cal R}}$, 
where ${\vec P}$ is the momentum of the particle and ${\vec {\cal R}}$ is its 
position vector. The constant $c$ was included to 
\vspace*{1.5cm}
%
match the dimensionalities, but in the numerical 
calculations we assume that $c=1$. Here we used a 
non-relativistic approximation, ${\vec P}=m{\vec v}$, 
where $m$ is the mass of the particle and $v$ is its velocity. We assumed a 
mass of $3200$ units, than we repeated our calculations by doubleing 
and halving it, and by using two other masses in this domain. 
This way we proved that the wavelength depends on the momentum, 
but not on the mass or velocity.
\begin{table}
\begin{tabular}{|c|c|c|c|c|} \hline 
$P \times 10^{-4}$  & 2   & 3  & 4  & 5  \\ \hline
$\Lambda$ & 0.40 & 0.27 & 0.22 & 0.16   \\
  \hline
\end{tabular}  
\caption{\it  Empirical study (moving particle) of the parameters: $\Lambda$ and $ P$}
\label{tab3}
\end{table}
We found that the wavelength $\Lambda$ of this 
wave depends on the momentum $P$.
We considered 
$D=40$ units, $d=1.5$ units and $\lambda=5 \times 10^{-3}$ units. 
We divided the screen into $70^2$ equal squares in order 
to do the numerical integration involved in reading the hologram. 

This calculation indeed suggests that within the limit of 
our errors the the Broglie relation, $\Lambda \sim 1/P$ 
is satisfied. The only thing remaining to be clarified is 
the origin of the spinorial structure for this wave. 
We propose that it has to be the same as that of the mapping wave. 

\section{Further Considerations}
We could ask the question: is there a more general argument that
tells us that quantum physics should come from an other principle?
In other way is there another argument to further 
justify, to strengthen our former reasoning 
of associating waves  with spinor structure to particles?  
The answer is positive, and we give this reasoning here, 
but its only disadvantage is that it is completely general, 
does not even give us the de Broglie relation nor does it tell us how to 
associate waves to particles, as we did in the former paragraphs. 

Variational principles can be applied to a wide variety of systems. 
These systems either have a finite degrees of freedom 
or infinite degrees of freedom. Another possibility does not exist.
The former case is the mechanics 
of a finite number of points and the latter case is basically a field theory.
In this section we discuss why the case of the point mechanics has 
difficulties  in the relativistic context, and propose a solution 
how this complication can be solved.

For a point like particle, in the non-relativistic context, the variational 
principle reads:

\begin{equation}
\int dt \left( m\frac{v^2}{2}-V \right) =minimum\, .
\label{1eq445}
\end{equation}

Here $m$ is the particle mass, $v$ is its velocity, 
and $V$ is its potential energy. This variational principle leads to 
the conservation of $H=\frac{1}{2} mv^2 +V$, which happens to be the energy.
In relativistic machanics, if we neglect interactions, there are 
two variational principles for a point-like particle. 
The first one is not manifestly 
covariant
\begin{equation}
- m c^2 \int \sqrt{1-\frac{v^2}{c^2}} dt =minimum\, .
\label{1eq446}
\end{equation} 

The function $H=\frac{mc^2}{\sqrt{1-\frac{v^2}{c^2}}}$ is the corresponding 
conserved quantity that is the energy as it was in the previous case. The only 
problem is that the formalism is not manifestly covariant.
The manifestly covariant principle reads: 

\begin{equation}
- m \int d\tau u^2  =minimum\, .
\label{1eq447}
\end{equation} 

But here $u$ stands for the four-velocity and $\tau$ for the proper time.
However, the conserved Hamiltonian associated to this 
action is zero. So, it is impossible to introduce for a point like particle 
a relativistic, manifestly covariant action principle that gives the right 
Hamiltonian. On the other hand, this is possible for fields.

How do we solve then the issue of point-like particles? One way would be
eliminating them completely, but that would be eliminating a part of 
reality, since such particles exist. An other way would be eliminating 
relativity or the need for a manifestly covariant description, which 
has the same disadvantage of ignoring reality. The only remaining possibility 
would be associating to 
any such a particle another system which does have a relativistic, 
manifestly covariant description, and that could be only a field. 
Any field is a function that depends on the spacetime, and that 
can be decomposed in plane waves, therefore it is a wave.
But the assertion that to every particle there is a 
corresponding wave is the basic assumption of quantum physics.
Therefore quantum physics, or at least a part of it, seems to follow 
from relativity. 

\section{Discussion \& Conclusions}
\label{S:conclusions}
Quantum theory predicted that regardless of the distance between the particles,  their polarizations would always be the same. The act of measuring one would force the polarization of the other. This shows a spooky exchange of informations between the two particles, which violates the principle of relativity. Alain Aspect's experiment \cite{aspect} at the university of Paris has shown that under certain circumstances subatomic particles such as electrons do communicate instantaneously. It doesn't matter whether they are 1 meter or 1 billion kilometers apart. This result is hard to digest, however a British physicist David Bohm \cite{d_bohm} claimed Aspect's findings imply that objective reality doesn't exist. Despite it's apparent solidity the universe at heart is phantasm, a gigantic and splendidly detailed hologram. 

In other words, the separateness of things is but an illusion, and all things are actually part of the same unbroken continuum. We see the separateness of things because we see the partial reality.
He gave a nice analogy of a fish in an aquarium, which is seen through two TV cameras, focused on at different angles, on two respective screens. It appears that there are two fishes, but simultaneously communicating, which is because either image of the fish contains only a partial information, i.e. a lower dimensional reality.

There are speculations that our universe could be a hologram of higher dimensional reality on the 4-D surface at its periphery. Theoretical results about black holes suggests that the universe could be like a gigantic hologram \cite{jacob}. So our innate perception that the world is 3+1 dimensional reality could be just an illusion, and our seemingly three dimensional universe can be a projection of a bigger reality. 

In terms of informations the holographic principle holds that the maximum entropy or information content of any region of space is defined not by its 'volume', but by its 'surface area'. 
According to John A. Wheeler of Princeton University the physical world is made of information, with energy and matter as incidentals.
Holography may provide a guiding torch to discover a better theory or 'Theory Of Everything'. So the final theory might be concerned not with fields, not even with spacetime, but rather with information exchange among physical processes
\cite{jacob}.

The revolution in the Holographic Principle is now a major focus of attention in many area of science e.g. gravitational research, quantum field theory and elementary particle physics. A popular account of holography can be found in \cite{susskind, susskind_1, taubes}. For a more technical discussion see \cite{tooft}.

In this work we discovered that the holographic projection of a point particle is a wave like structure.
We have also noted that for a fixed distance (D) of particle from the screen of hologram and fixed wavelength of the reference \& object beam the there exists a relation between momentum and wavelength of the wave pattern associated with the particle. This relation is similar to the famous de Broglie-type relation: $\Lambda \sim 1/p_{rms}$. 

Now, we conjecture here that since the relation (or the proportionality) between wavelength and momentum is similar to that of the de Broglie relation, it could be the the wave-pattern which we associate with all the particles in quantum mechanics. If this is so, we then understand the origin of the wave like nature of quantum particles. This will also show a holographic basis for the existence of this intire physical world. 
Whatever we see around is a holographic projection of a bigger reality. In other words, that a point particle is not seen as a point, but a fuzzy Gaussian wave that has been observed in quantum mechanical experiments. We can dare to say that the holographic projection wave of the point particle is essentially the so-called de Broglie wave. 

We also noted that there is a need to give a physical significance to the grid
points of the hologram screen, which prove another assumption of quantum 
gravity: the space is quantised.   
We want to go one step further and argue that for the macro particles (classical objects) this fuzziness, noise and wave pattern due to holographic projection are weak and so hard to observe. To test this argument one should bear in mind that now each point of the particle is a source of object wave which will interfere with the reference wave and with themselves. In our investigation, we got an impression \cite{prep} that although the computation is much more complicated, but the final reconstructed image of the object go sharper as the size of the test object grows up. Such that for the classical objects the fuzziness due to the holographic process will almost disappear and we can see only the physical size and shape of the particle. 

As said before the relation $\Lambda \sim 1/P$ and the wave pattern we 
found as a holographic projection of a point particle exist only in certain 
domains of the parameters that we used. It would be interesting to investigate 
the domains of these parameters with some boundary conditions e.g. 
$\lambda \ll d$, $\lambda \ll D$, $\lambda \ll 1/N$, $d \ll D$ and 
$\Lambda > d/N$. 
Even in the real world there is a domain of validity 
for every physical theory. For example, quantum mechanics and 
field theory are valid if the calculated de Broglie wavelength 
is much bigger than the planck scale, which can be perceived as 
the minimal disatnce between two points in the quantized space-time 
continuum. And likewise, this wavelength must be much smaller than 
the astronomical distances where general relativity sets in.
In our next work we want to investigate the validity domain of 
our conclusions if and whether or not this domain coincides with the 
domain dictated by the former criteria \cite{prep}. 
       
\section{Acknowledgments}
The work of BCC was supported by Funda\c{c}\~{a}o para a Ci\^{e}ncia e a Tecnologia through the grant SFRH/BPD/5719/2001.

\end{document}